\documentclass[12pt,reqno]{amsart}

\usepackage{amsmath}

\newtheorem{thm}{Theorem}
\newtheorem{prop}[thm]{Proposition}

\makeatletter \@addtoreset{equation}{section} \makeatother

\newcommand{\ben}{\begin{enumerate} }
\newcommand {\een}{\end{enumerate} }
\newcommand{\W}{\mbox{${\mathfrak W}$}}
\newcommand{\wat}[1]{\mbox{$\mathfrak{W}(\mathfrak{A}^{#1})$}}

\newcommand{\ehat}{\mbox{$\widehat{\eta}^{t}_{s}$}}
\newcommand{\enohat}{\mbox{$\eta^{t}_{s}$}}
\newcommand{\winf}{\mbox{${\mathfrak W}^{\infty}$}}
\newcommand{\states}{\mbox{${\mathcal K}\winf$\hspace{1mm}}}

\newcommand{\dirsys}{\mbox{$\{\wat{t}, \ehat, {\mathbf J}\}$ }}
\newcommand{\threads}{\mbox{$E_{\infty}$}}
\newcommand{\invsys}{\mbox{$\{E_t, \enohat, {\mathbf J}\}$ }}
\newcommand{\dirlim}{\mbox{$\{\winf, \sigma_t, {\mathbf J}\}$ }}
\newcommand{\invlim}{\mbox{$\{E_{\infty}, \rho_t, {\mathbf J}\}$ }}
\newcommand{\net}{\mbox{$(\mu_t)_{t \in {\mathbf J}}$}\hspace{1mm}}
\newcommand{\C}{\mbox{${\mathcal C}(X)$}}
\newcommand{\Co}{\mbox{${\mathcal C}(X_K)$}}

\newcommand{\glosta}{\mbox{$\mathcal{K}\C$}}
\newcommand{\R}{\mbox{$\mathbf{\sf R}$}}
\newcommand{\zemu}{\mbox{$\zeta_{\mu}$}}
\newcommand{\B}{\mbox{$\mathcal{B}$}}
\newcommand{\Bo}{\mbox{${\mathcal B}(X) $}}

\newcommand{\chip}[1]{\mbox{$\chi^{(X)}_{#1}$}}
\newcommand{\chipo}[1]{\mbox{$\chi^{(\Omega)}_{#1}$}}
\newcommand{\Q}{\mbox{${\mathfrak P}$ }}

\newcommand{\wo}{\mbox{${\mathfrak W}$}}
\newcommand{\Ws}{\mbox{${\mathcal{K}}{\mathfrak W}$}}
\newcommand{\spg}{\vspace{5mm} \noindent}
\newcommand{\hk}{\mbox{Haag-Kastler }}
\newcommand{\prf}{\mbox{{\em Proof.\hspace{2mm}}}}
\newcommand{\qd}{\mbox{\hspace{5mm}\rule{2.4mm}{2.4mm}}}
\newcommand{\cc}{compact convex }

\newcommand{\So}{{\mathcal S}}
\newcommand{\ie}{{\em i.e.,} }
\newcommand{\beq}{\begin{quote}
  }
\newcommand{\eq}{\end{quote}}

\begin{document}

\begin{center}
{\huge {\bf An algebraic theory of infinite classical lattices III:  Theory of single
measurements}}

\vspace{20mm} {\bf Don Ridgeway}

\vspace{15mm}
Department of Statistics,\\
North Carolina State University,\\
Raleigh, NC  27695\\
ridgeway@stat.ncsu.edu
\end{center}

\vspace{1cm} \hspace{53mm} {\bf Abstract}

\vspace{5mm} {\small This  is the third in a series of papers  dealing with the algebraic
theory of infinite classical lattices. This paper presents a theory of single
measurements on the lattice which we represent as comprising a finite subvolume---the
system of measurement---immersed in an infinite surround or ``heat bath'' which
determines the system's  state. We consider the class of all stationary distributions on
the set of microcanonical states of the infinite lattice.   The theory addresses the
question, ``For a lattice initially in state A, say, what is the probability that
measurement of a certain quantity will take a value in (a,b)?''} Discussion includes
description of the source of randomness in a measurement as well as characterization of
the given states A.

\vspace{1cm} \noindent MSC 46A13 (primary) 46M40 (secondary)

\newpage

\vspace{5mm} \noindent{\bf {\large I \hspace{4mm} Introduction }}

This is the third in a series of papers presenting an algebraic theory of  measurements
on classical infinite lattices. The first paper \cite{ridga} dealt with construction of
Segal algebras\cite{sega47} for lattices.  It was applied to groups of symmetries and
symmetry breakdown and to equilibrium Gibbs states. Construction used the classical
analogue of the \hk axioms \cite{haag64} from quantum field theory (QFT). The second
paper \cite{ridgb} gave an axiomatic theory of measurement on such systems, making use of
Mackey's axioms for a quantum theory \cite{mack63}. The present paper deals with the
theory of local measurements on an infinite lattice.

The algebraic observables lead to questions of the following form:  ``For a measurement
on a system initially in a given state A, say, what is the probability the outcome will
lie in the interval $(a,b)$ for arbitrary $a < b$?'' This will bring together results
from the general theory of the first two papers that bear on single measurements. It will
include description of the source of randomness in measurements as well as
characterization of the given states A. We give sufficient detail to obtain a substantial
picture of the algebraic construction, but without proofs.

\vspace{5mm}

\vspace{5mm} \noindent{\bf {\large II \hspace{4mm} Measurements on an infinite lattice.
}}

\begin{center} {\bf A. Lattice structure.} \end{center}

The lattices are infinite arrays of sites of one or more dimensions, indexed by the set
$\mathbf{T}$. If $\Omega_o$ denotes the set of possible configurations on an individual
site, then  the configuration of the whole lattice is the product $\Omega = \mathbf{P}_{i
\in \mathbf{T}} \Omega_i$, where $\Omega_i = \Omega_o$ for all $i \in \mathbf{T}.$ The
instantaneous configuration of the lattice is therefore a point $(\omega_i)_{i \in
\mathbf{T}} \in \Omega$, giving the configuration at each site at that instant. By
design, algebras constructed from this description contain no detailed information about
the lattices involved. We shall see below that it is possible to specialize to a
particular lattice structure at the level of algebraic states themselves through
application of  expectation values of the microcanonical (MC) states obtained from TL
calculations for that structure.

\begin{center}{\bf  B. Local measurements.} \end{center}

For the description of a measurement here, we will treat the infinite lattice as
comprising a finite system immersed in its infinite surround---a generalized
``temperature bath''---taking as possible systems of measurement the finite subvolumes of
the lattice. Denote the set of all finite systems by $\mathcal{P} = (\Lambda_t)_{t \in
\mathbf{J}}$, where $\mathbf{J}$ is an index set partially ordered by inclusion: $s \leq
t$ iff $\Lambda_s \subseteq \Lambda_t$.

The local observables are defined to express a convention from traditional theory of
measurements. In statistical thermodynamics, the values used in the Gibbs ensembles, for
example, for the intensive variables of exchange are their values as measured {\em in the
surround}. Thus, for systems that can exchange only heat, Guggenheim writes, ``$\beta$
$[=1/kT]$ is determined entirely by the temperature bath {\em and so may be regarded as a
temperature scale}'' (\cite{gugg57}, p.65). A similar rule obtains for the pressure and
other intensive variables, for the same reasons. Hence, the state of the system is
determined by the bath.

This convention has been incorporated into the definition of the local observable spaces
here by choice of a class of functions called  {\em functions from the outside} to
represent measurements on any given system $\Lambda_t$. The class of functions appears
already in the theory of Gibbs states, where it supplies the theory's observables
(\cite{pres76},  \cite{ruel78}). Functions from the outside are defined as functions on
the whole configuration space $\Omega$ with values that depend only on configurations
{\em outside} a system $\Lambda_t$. Note that with this definition of  observables, the
systems of measurement themselves do not depend on walls or containers introduced into
the lattice for their definition.

For each $t \in \mathbf{J}$, we define the set $\wat{t}$ of all bounded Borel-measurable
functions of this form. The $\wat{t}$ are partially ordered by inclusion. Since every
measurement from outside the system $\Lambda_t$ is obviously a measurement from outside
$\Lambda_s \subset \Lambda_t$, $\wat{s} \supseteq \wat{t}$ for all $s \leq t$. The
partial order $\wat{s} \leq \wat{t}$ iff $\wat{s} \supseteq \wat{t}$ gives $s \leq t
\Rightarrow \wat{s} \leq \wat{t}.$ In the \hk terminology, the set of pairs $(\Lambda_t,
\wat{t})_{t \in {\mathbf J}}$ forms the {\em texture} of the lattice. All of local theory
is in terms of this texture.

Measurements on different parts of the lattice will be related  by postulate as follows.
For all nested pairs of systems $\Lambda_s, \Lambda_t$ with $\Lambda _s \subset
\Lambda_t$, it is assumed that there exists a morphism $\ehat:\wat{s} \rightarrow
\wat{t}$ that maps any measurement $f^s \in \wat{s}$ on $\Lambda_s$ to the measurement
$\ehat f^s \in \wat{t}$ of the same quantity on $\Lambda_t$.

\begin{center} {\bf C. The measurement.} \end{center}

We consider the class of measurements described as follows. The lattice is initially
prepared at internal equilibrium in a given stationary state. At the beginning of the
measurement, it is suddenly isolated so as to freeze it in a MC state randomly chosen
from the ensemble defined by the prepared state. The value of a measurement is its
expectation value on a lattice in this MC state, as obtained, for example, by a
thermodynamic-limit (TL) calculation. Instantaneous operations are not new to CSM, since
the phase function itself is instantaneous.  We shall show that measurements on the MC
states, and on these alone,  have zero variance.

\vspace{5mm} \noindent{\bf {\large III \hspace{4mm} Construction of Algebras}}

\begin{center}  {\bf A. The direct limit }$\dirlim = \lim^{\rightarrow} \dirsys$. \end{center}

As already stated, the construction of a Segal algebra from a local texture applied the
classical analogue of the axioms of Haag and Kastler \cite{haag64} for a quantum field
theory (QFT). The algebraic theory removes the distinction of particular systems within
the lattice. Already the first step in the construction defines a space $\winf$ with
elements representing equivalence classes of the local measurements. $\winf$ is the
Banach-space direct limit of the $(\wat{t})$ and their morphisms $(\ehat)$. Measurements
$f^s \in \wat{s}$ and $f^t \in \wat{t}$ are identified with the same element $[f] \in
\winf$ if, and only if, they measure the same quantity on their respective systems.   For
each $t \in {\mathbf J}$, the morphisms $\sigma_t: \wat{t} \rightarrow \winf$ map  local
observables to their respective equivalence classes in $\winf$.

\begin{center}{\bf  B. States.} \end{center}

In the theory of Banach spaces, states are positive linear operators of norm 1. They are
are well-defined on both kinds of Banach spaces we have named, $\wat{t}$ and $\winf.$ We
denote the states on $\wat{t}$ by $E_t$, for all $t \in {\mathbf J}$, and use the
category-theoretical functor ${\mathcal K}$ for the limit space, writing $\mathcal K
\winf.$ We shall later have the Banach spaces $\W$ and $\C$ with their states $\Ws$ and
$\glosta$.

\begin{center} {\bf C. The index  set} $\invlim = \lim^{\leftarrow} \invsys.$
\end{center}

Application of the algebraic theory to local measurements depends on a second
category-theoretical limit involving the local states $(E_t)$. The sets of states are
compact convex sets in unit ball of the duals of the $(\wat{t}$. We define the following
set of morphisms. For all $s < t$, define $\enohat :E_t \rightarrow E_s$ as the canonical
mapping $\enohat \mu_t(f^s) = \mu_t(\ehat f^s)$. The inverse-limit limit object for the
\cc sets $(E_t)$ is the set $\threads = \{(\mu_{t})_{t \in \mathbf{J}} \in
{\mbox{{\bf{\sf P}}}}{}_{t \in {\mathbf J}} E_{t} : \mu_{s}= \enohat
\mu_t\hspace{1em}\forall s \leq t,\hspace{.5em}s,t \in \mathbf {J}\}$. The morphisms
$\rho_t: \threads \rightarrow E_t$ give the component state for the system $\Lambda_t$,
for all $t \in {\mathbf J}.$ In a crucial result (\cite{ridga}, Theorem V.5) it is shown
that there is a unique 1:1 correspondence $\mu \leftrightarrow \phi_{\mu}$ between the
elements of $\threads$ and those of the set of states $\states$. This enables us to index
$\states$ with the corresponding elements of $\threads$, writing $\phi_{\mu} \in
\states$.

The elements of $\threads$ are called {\em threads}. They are nets of local states from
each finite system, written $\net$, and specifically those which satisfy the homogeneity
condition $\mu_{s}(f^s) = \enohat \mu_t(f^s) = \mu_t(\ehat f^s)$ for all $s < t$.  Recall
that $\ehat f^s \in \wat{t}$ is the local observable that measures the same physical
quantity on system $\Lambda_t$ as $f^s$ measures on $\Lambda_s$, for all $s < t$. As we
shall later show, this is the condition that measurements of the same quantity on nested
systems have the same expectation values. Each thread is therefore an atlas of local
states for a lattice internally at equilibrium.

\begin{center} {\bf D. The Segal algebra} \W. \end{center}

Reference \cite{ridga} was a study of the properties of compact convex sets of $\states$.
The transformation of the Banach space $\winf$  to algebraic observables   took two
steps. It was shown that for any fixed \cc subset $K \subset \states$, one may construct
an $MI$-space $\wo_K$, {\em i.e.}, a Banach lattice with unit, as  an order-unit
completion of $\winf$. From the structure theorem for $MI$-spaces, $\W_K$ may then be
represented be the algebra $\Co$ of continuous functions on a certain compact space $X_K$
(\cite{sema71},Theorem 13.2.3). The elements of $\Co$ are the {\em algebraic observables}
of the theory of $K$. It was shown that (a) the set of states $\mathcal{K}\Co$ is
isomorphic with $K$ itself and (b) $X_K$ is homeomorphic with the set
$\partial_e\mathcal{K}\Co$ of extremal points of $\mathcal{K}\Co$.

The theory of measurement has to do with a particular choice of $K$, the set of
stationary distributions on the lattice. Let $E \subset \winf$ be the set of all MC
states on the lattice, and let $K$ be the closed convex hull $\overline{\mbox{co}}(E)$ of
$E$. By MC states, we mean those states in $\winf$ identified with a thread $\mu = \net
\in \threads$ with  projections of a given MC state as components. The MC states are
specified by pairs of values of the energy and particle-number densities.  Since all
stationary distributions are written as Borel functions of these two constants, they may
be regarded as distributions over the set of MC states. It is readily shown that
$\partial_eK = E$, \ie that our set $X_K$ is exactly the set of MC states on the lattice.
Since interest here is specifically in this choice of $K$, we henceforth suppress the
subscript $K$ and simply write $\W$, $\C$, and $\glosta$.

The structures $\C, \glosta$ complete the construction of a Segal algebra and its states
from the texture of an underlying infinite lattice.  The treatment of the infinite
lattice by algebraic theory is noteworthy.  There is nothing comparable  to an approach
to infinite size such as one finds, in particular, in the TL calculation. The observables
are defined from the outset for an infinite lattice in terms of its topology, and the
local states are simply the nonempty set of positive linear elements of norm 1 in the
duals of the Banach spaces $(\wat{t})_{t \in {\mathbf J}}$. The direct and inverse limits
of the infinite nets $(\wat{t})$ and $(E_t)$ from the category theory always exist.

\vspace{5mm} \noindent{\bf {\large IV \hspace{4mm} Statistics of local measurements}}

\begin{center} {\bf A. Expectation values.} \end{center}

The  Segal structure thus obtained allows us to write a probability theory for the
measurements. Its construction must contain the means  of translating those results back
to laboratory-scale measurements and the local observables.

Up to this point, nothing has been given to impart a probabilistic meaning to the
theory's states. The properties of the space $\C$, $X$ compact, remedy this situation. In
fact, the Riesz Representation Theorem for these spaces provides a 1:1 correspondence of
the states on $\mathcal{K}\C$ and the set of all probability distributions in $X$. Thus,
for any state $\zemu \in \glosta$, there exists a unique Radon probability measure
$\sigma_{\mu}$ on $X$ such that

\[ \zemu(f) = \int_X f(x) d\sigma_{\mu}(x)  \hspace{5mm}\forall f \in \C, \mu \in
\threads \hspace{15mm} (1) \]

\spg Since this is a decomposition of the state $\sigma_{\mu}$ into extremal states, we
should note that a Choquet decomposition theorem \cite{phel66} was needed  in the
construction for the uniqueness in the indexing by $\threads$ and was proven in Reference
(\cite{ridga} (Theorem III.19). as such for $\states$ and its compact convex subsets $K$.

By hypothesis, the {\em lattice states} are the  stationary probability distributions on
the set of MC states of the lattice. By eq.(1), it is proper to regard the state
$\zeta_{\mu}$ as the corresponding {\em expectation-value operator} on the space $\C$ of
observables. As a first application, note that the extremal states are exactly the
multiplicative states, {\em i.e.,}, those for which $\zemu (f^2) = \zemu (f) \cdot
\zemu(f) = \zemu(f)^2$. Hence, for these states, and only for these, the variance of the
measurement $f$ is 0. Eq.(1) has the classical form of an expectation value of the
observable as an integral over the algebraic ``phase space'' $X$. Thus, the
transformation from the Segal algebra $\W$ to $\C$ is  the classical analogue of the
Gelfand-Naimark-Segal (GNS) construction in QFT.

The translation of algebraic results into terms of local measurements is based on the
relation

\[ \mu_t(f^t) = \zemu(f) \]

\spg (\cite{ridga}, eg.4.6), where $f \in \C$ is the representation in $\C$ of the
equivalence class $\sigma_t f^t = [f] \in \winf$. Rewriting of eq.(1) yields

\[ \mu_t(f^t) = \int_X f(x) d\sigma_{\mu}(x)  \hspace{5mm}\forall f^t \in \wat{t}, \mu_t \in
E_t, t \in {\mathbf J} \hspace{15mm} (3) \]

\spg The construction has completed its task in this equation. The transformation   to
algebraic observables converts the local problem with its observables $f^t \in \wat{t}$,
for any finite system $\Lambda_t$, to an integral over the set of MC states with respect
to a given state $\sigma_{\mu}$, with its evident statistical implications.

\begin{center} {\bf B.The Mackey Theory.} \end{center}

In order for the classical  Segal algebra $\C$ to satisfy the first six of Mackey's
axioms for a quantum theory, its states and the space $X$ must satisfy certain
conditions. We give them as follows. Regarding states, define the set of restrictions
$\So = \{\zemu |_{\Q}, \mu \in \threads \}$. We can show, first, that if for any two
given sets $E, F, \in \Bo$ $\zemu(\chip{E}) \leq \zemu(\chip{F})$ $\forall \mu \in
\threads$, then $E \subseteq F$. That is, the set $\So$ is {\em full}.  Furthermore, if
$(t_n) \in [0,1]$, $\sum_n t_n = 1$, and $(\zeta_{\mu_n}) \in \So$, then $\sum
_1^{\infty} t_n \zeta_{\mu_n} \in \So$. Hence, $\So$ is {\em strongly convex}. These are
the two requirements on $\So$. The phase space of the algebraic theory is the phase space
$X$. The condition on it concerns its topology.  The axioms need that $X$ be Stonean, \ie
compact and extremely disconnected (e.d.). It is shown, in fact,  that all open sets in
$X$ are clopen (closed-and-open), so that $X$ is always Stonean,   without any conditions
on the underlying lattice (\cite{ridga}, Theorem V.3) .

The requirement of a Stonean phase space is a stringent condition that excludes most
physically interesting phase spaces. In particular.  the lattice configuration space
$\Omega$ is not e.d., so that the space $C(\Omega)$ of continuous function---the
observable space in the TL calculations \cite{ruel78}--- is not  a suitable set of
observables for the present theory of measurement. For lattices, the space   $\Omega =
\mathbf{P}_{t \in \mathbf{J}} \Omega_t$ is  a compact totally disconnected
(0-dimensional) space which is not e.d. This is shown as follows. The Cartesian product
is totally disconnected compact because each $\Omega_t$ is (Tychonoff's Theorem).
However, if two spaces $X$ and $Y$ are compact and the product $X \times Y$ is e.d., then
one of these factors is finite and the other e.d.(\cite{sema71}, Note, 24.2.12). Clearly
we can write $\Omega$ as the product of two infinite compact spaces.

Denote the set of idempotents of $\C$ by $\Q$.  The idempotents play a special role in
the theory. They are easily characterized. For any $F \subset X$, denote by $\chip{F} : X
\rightarrow \{0,1\}$ the {\em characteristic function}, \ie $\chip{F}(x) = 1$ if $x \in
F$ and 0 otherwise. Since all sets in $\Bo$ are clopen---and here one sees the Stonean
requirement---$\chip{F} \in \C$ for all $F \in \Bo$. The characteristic functions are
exactly the idempotents of $\C$, since for any $F \in \Bo$, $(\chip{F} \cdot \chip{F})(x)
= \chip{F}(x)\cdot\chip{F}(x) = (\chip{F}(x))^2 = \chip{F}(x)$.

$\Q$ is a complete Boolean algebra with the lattice operations $\chip{E} \vee \chip{F} =
\chip{E \cup F}, \chip{E} \wedge \chip{F} = \chip{E \cap F}$, and the complementation
$(\chip{F})^{\prime} = 1 - \chip{F} = \chip{F^{\prime}}$. The completeness of $\Q$ means
that the infinite operations $\vee_n \chip{F_n} = \chip{\cup_nF_n}$ and $\wedge_n
\chip{F_n} = \chip{\cap_nF_n}$ exist. This is equivalent to the condition that $X$ is
Stonean (\cite{port88}, Theorem 6.2d). The mapping $\phi : \Q \rightarrow \Bo$ defined
by $\phi(\chip{F}) = F$ is a lattice isomorphism from the class $\Q$ of idempotents of
$\C$ onto the topology $\Bo$ of $X$.

As continuous functions of $X$, all idempotents are integrable. For any Borel set $B \in
\B$ of the real line, the probability of the extremal states for which $f(x) \in B$ is

\[ \zeta_{\mu}(\chip{[f \in B]}) = \int_{X} \chip{[f \in B]} d\sigma_{\mu}(x) =
\int_{ [f \in B]} d\sigma_{\mu}(x) \hspace{15mm} (4)         \]

\spg for any $\chip{[f \in B]} \in \C$. This is exactly the probability $\sigma _{\mu}([f
\in B])$ of drawing a state $x$ at which $f(x) \in B$ at random from $X$ when the lattice
is in state $\zemu$ or, equivalently, when the probability distribution on $X$ is
$\sigma_{\mu}$.

The purpose of the axioms is to turn eq.(4) into a probability law. Let $(\R,
\mathcal{B})$ be the real line with its Borel topology. The function $Q:{\mathcal B}
\rightarrow \Q$ is called a  {\em $\Q$-valued measure} on $\R$ iff the following obtain:

(a) $ Q(\emptyset) = 0,$ $ Q(${\bf {\sf R}}) = 1;

(b) If $(B_n)$ is any family in ${\mathcal B}$, and $B_i \cap B_j =
\emptyset$ for all $i \neq j$, then $Q(\cup B_n) =$

\begin{tabbing}
12345\=\kill \> $\bigvee Q(B_n)$.
\end{tabbing}

\noindent Let $\mathcal{O}$ be the set of all $\Q$-valued measures on $\R$. For any $f
\in \C$, define the measure $Q^f : \B \rightarrow \Q$ by $Q^f(B) = \chip{[f \in B]}$,
where $[f \in B]$ is the preimage $f^{\leftarrow}(B)$. Then $Q^f$ is a lattice
homomorphism on $\B$ into $\Q$.   By Mackey's Axiom VI, for every $Q \in \mathcal{O}$,
there exists an $f \in \C$ such that $Q = Q^f$. Hence, $\C = \mathcal{O}$ is an
isomorphism. By Axiom I, the relation $\zemu(Q^f(.)): \B \rightarrow [0,1]$ is a
probability law on $(\R,\B)$ for any $f \in \C$ and any $\zemu \in \glosta.$ By Axiom II,
states separate observables, and observables separate states.  That is, if $f \neq g$,
then there exists a state $\zemu \in \glosta$ such that $\zemu(f) \neq \zemu(g)$; and if
$\zemu \neq \zeta_{\nu}$, there exist an observable $f \in \C$ such that $\zemu(f) \neq
\zeta_{\nu}(f)$.

The construction of the algebraic observables from the local texture took three steps.
The first was application of the morphisms mapping local observables to their equivalence
class in $\winf$, \ie $\sigma_t : \wat{t} \rightarrow \winf$ sending $f^t \mapsto [f].$
The second used the Kadison transformation \cite{kadi51} $\Delta_K : \winf \rightarrow
\W_K$.  $\Delta_K$ is an order-preserving isometry mapping $\winf$ onto a dense subset of
$\W_K$ and sending $[f] \mapsto \hat{f}$, say. (We restore the subscripts $K$ for the
moment to indicate the choice of the compact convex subset $K \subseteq \states$ in the
construction.) Finally, the third used the representation of the $MI$-space $\W_K$ as
$\C$ by an affine isometric isomorphism $\psi_K:\W_K \rightarrow \Co$, mapping $\hat{f}
\mapsto f$ (\cite{sema71} Theorem 13.2.3). For each $t \in {\mathbf J}$, we define the
compose $\gamma_t = \psi_K \circ \Delta_K \circ \sigma_t :\wat{t} \rightarrow \Co$. Then
$\gamma_tf^t = f$.

Application of the algebraic observables to represent  local measurements requires the
following result.

\begin{prop} For any $F \subset X$ and  $f \in \C$, denote $f(F) = A_F.$ Let $f^t \in \wat{t}$ be
any local observable  such that $\gamma_tf^t = f$.  Then $\gamma_t\chipo{[f^t \in A_F]} =
\chip{[f \in A_F]}$. \end{prop}

\noindent \prf Fix any $x_{\mu} \in F$. Then $f(x_{\mu}) = x_{\mu}(f) \in A_F$, and
therefore $\mu_t(f^t) = f^t(a ) \in A_F$ or $a  \in [f^t \in A_F]$ for some $a  \in
\Omega$, since $x_{\mu}$ is extremal. Hence, $\chipo{[f^t \in A_F]}(a ) =
\mu_t(\chipo{[f^t \in A_F]}) = 1$, and therefore $x_{\mu} \in F$ iff $\mu_t(\chipo{[f^t
\in A_F]}) = 1$. But $\gamma_tf^t(x_{\mu}) = x_{\mu}(\gamma_tf^t) = \mu_t(f^t)$ for all
$\mu \in \threads$, $f^t \in \wat{t},$ so that in particular, $\gamma_t\chipo{[f^t \in
A_F]}(x_{\mu}) = \mu_t(\chipo{[f^t \in A_F]}) = 1$ iff $x_{\mu} \in F$. But
$\chip{F}(x_{\mu}) = 1$ iff $x_{\mu} \in F$.  \qd

\spg The relation $\mu_t(f^t) = \zemu(f)$ takes us from local to algebraic theories, with
$f = \gamma_tf^t$.  However, this is a purely formal relation without Proposition l. With
the proposition, it becomes

\[ \mu_t(\chipo{[f^t \in B]}) = \zemu(\chip{[f \in B]}) \hspace{4cm} (4) \]

\spg The probability of finding the value of the local observable $f^t$ in the set $B
\subset \R$, given the initial state $\zemu$, is the same as that of finding the value of
its algebraic image $\gamma_tf^t$ in $B$.

Note that the expectation value on the left would require the traditional integration
over phase space with respect to a distribution on $\Omega$. Mackey states the problem
with this class of integrations as follows: ``Since a point in phase space has no
physical meaning, neither does the notion of a probability measure in phase space''
(\cite{mack63}, p.62). What Mackey intends with ``physical meaning'' has been constant in
axiomatic theory since it was stated by Birkhoff and von Neumann:

\beq Before a phase-space can become imbued with reality, its elements and subsets must
be correlated in some way with experimental propositions'' , {\em i.e.,} with the Borel
sets of the real line $\R$ and its products $\R^n$ (\cite {birk36}, p.825).\eq

\spg The compose $\phi \circ Q^f$ mapping $B \mapsto [f \in B]$ is a lattice homomorphism
on the Borel sets $\B$ of $\R$ into the Borel sets $\B(X)$ of $X$. That is, in the
language of Birkhoff and von Neumann, it is just such a {\em correlation} of the two
algebras, for every choice of observable $f \in \C$.

In general, comparisons with the quantum theory are facilitated by choosing the set $\Q$
directly as the phase space of the theory (rather than $\Bo$), as Mackey describes, with
the homomorphisms $Q^f$ themselves as the correlations.  As already discussed, $\Q$ is a
complete Boolean algebra. Its elements are the idempotents of $\C$, comparable to the
projectors on Hilbert space. The comparison is sharpened  by a classical spectral theorem
(\cite{ridgb}, Proposition IV.2) which says that every observable $f \in \C$ has a unique
{\em spectral decomposition} $f = \int_{-\infty}^{\infty} \lambda dQ^f(\lambda)$, where
$Q^f(\lambda) = \chip{[f \leq \lambda]} \in \Q$.

\spg {\sc Acknowledgement}. The author wishes to express his gratitude to Rudolf Haag for
his many suggestions during the writing the papers of this series.

\end{document}